# *In situ* XRD Study of Strain Evolution in AlGaN/GaN HEMT at High-Temperatures up to 1000 °C

Botong Li[1, †, *], Bobby G. Duersch[2, †], Hunter Ellis[1], Imteaz Rahaman[1], Aidan Belanger[3], Zlatan Aksamija[3], Brian Roy Van Devener[2], Kathy Anderson[4], Kai Fu[1, *]

[1]*Department of Electrical and Computer Engineering, The University of Utah, Salt Lake City, UT 84112, USA*

[2]*Utah Nanofab Electron Microscopy and Surface Analysis Laboratory, The University of Utah, Salt Lake City, UT 84112, USA*

[3]*Department of Materials Science and Engineering, The University of Utah, Salt Lake City, Utah 84112, USA*

[4]*Utah Nanofab, Price College of Engineering, The University of Utah, Salt Lake City, UT 84112, USA*

[*]Corresponding author: Botong Li (E-mail: u1540316@utah.edu); Kai Fu (E-mail: kai.fu@utah.edu)

[†]Authors contributed equally

**Abstract:** The thermal stability and structural evolution of a GaN high-electron-mobility transistor (HEMT) heterostructure grown on a Si (111) substrate were investigated using *in situ* high-temperature X-ray diffraction (HT-XRD), reciprocal space mapping (RSM), Raman spectroscopy, and rocking-curve (RC) analysis at varying temperatures. The heterostructure, consisting of a p-GaN cap, an AlGaN barrier, and a GaN channel supported by two AlGaN/AlGaN superlattice (SL) buffer layers, maintained clear and periodic satellite peaks up to a temperature of 1000 °C, confirming excellent structural integrity. Symmetric and asymmetric RSM results reveal that both the Si and GaN diffraction peaks left shift with increasing temperature, consistent with thermal expansion, and show no significant broadening or relaxation throughout the heating process. The *c*-lattice constant follows the theoretical expansion predicted by the multi-frequency Einstein model, whereas the *a*-lattice expansion is slower due to in-plane strain constraints imposed by the underlying Si substrate and buffer layers. Rapid lattice contraction during the fast-cooling stage induces a residual compressive strain of roughly 0.3% in the GaN channel after cooling. Raman spectra further confirm this strain state through a blue shift (~1.5 cm$^{-1}$) of the GaN E$_2$ (high) phonon mode, corresponding to an in-plane strain of 0.2%. Rocking-curve analysis revealed an increase in both screw and edge dislocation densities by 28% and 12%, respectively. These results collectively demonstrate that the GaN HEMT heterostructure exhibits robust crystalline stability up to 1000 °C, with only minor strain redistribution and limited dislocation activity, providing experiment evidence for GaN devices' applications under high-temperature condition.

Gallium nitride (GaN) high-electron-mobility transistors (HEMTs) have emerged as one of the most promising device architectures for next-generation high-frequency and high-power electronic systems[1-3]. Owing to their wide bandgap (3.4 eV), high electron saturation velocity (~2.5×10$^7$ cm/s), and superior breakdown field (~3.5 MV/cm)[4-6], GaN-based HEMTs are extensively applied in radio-frequency (RF) power amplifiers[7, 8] and power conversion modules that require high efficiency and compact form factors[9-11]. At present, GaN devices can operate reliably under elevated temperatures, highlighting the intrinsic potential for high-temperature electronics[12-14]. However, the present high-power GaN technology is still mainly optimized for operation below ~500 °C, and the device behavior under more extreme thermal conditions remains insufficiently understood[15-18]. At such temperatures, several coupled degradation mechanisms become critical, such as the large lattice and thermal-expansion mismatch in GaN-on-Si structures[19-21]; the interdiffusion, agglomeration, or phase transformation of metal contacts[22, 23]; and the strain redistribution within the HEMT heterostructure[24-26]. These effects underscore the need to clarify how strain, contact integrity, and electrical performance co-evolve when devices experience extreme thermal loads. To date, the high-temperature reliability of GaN HEMTs has been mainly investigated through post-annealing or ex situ characterization[27-32]. Direct *in situ*, temperature-dependent measurements of lattice behavior in GaN-on-Si HEMTs remain scarce due to the experimental challenges of maintaining structural integrity and measurement precision during heating.

In this work, we present a comprehensive *in situ* crystal structure study of an E-mode AlGaN/GaN HEMT heterostructure using high-temperature X-ray diffraction (HT-XRD), reciprocal space mapping (RSM), and complementary Raman spectroscopy. The combination of symmetric and asymmetric RSM measurements enables simultaneous monitoring of both vertical and lateral lattice responses up to 1000 °C, while Raman analysis provides independent strain verification. In addition, rocking-curve measurements before and after thermal cycling are used to quantify the evolution of screw and edge dislocation densities. This systematic approach provides direct insight into the lattice stability and defect dynamics of GaN-on-Si HEMTs under extreme thermal conditions, offering valuable guidance for the design and reliability optimization of high-temperature GaN power devices.

Crystal structure was evaluated by *in situ* HT-XRD, using a Brüker Discover D8 XRD diffractometer, an Eiger 2R 250k detector, and an Anton Paar DHS 1100 heating stage. The HEMT sample was enclosed by a graphite dome during the HT-XRD process to both protect the instrumentation and maintain the heat uniform. To evaluate the thermal stability and crystalline quality of the HEMT heterostructure, *in situ* HT-XRD measurements were conducted *2θ–ω* scans and reciprocal space mapping (RSM) in air at temperatures of 27 °C, 500 °C, 600 °C, 700 °C, 800 °C, 900 °C, and 1000 °C, followed by a final measurement at 27 °C. The temperature ramping rate between different setpoints was set to 300 °C min$^{-1}$ with a delay time of 5 minutes for the sample to reach equilibrium. At each prescribed temperature, a measurement duration was approximately 30 minutes (this included alignment, 2θ-ω scans, rocking curves, and RSMs). With the graphite dome still secured, dry compressed air was used to cool the sample from 1000 °C to 27 °C, taking roughly 15 minutes.

Figure 1(a) illustrates the overall layer configuration of the E-mode HEMT structure, consisting of a p-GaN cap layer, an AlGaN/AlN barrier, and a GaN channel with thicknesses of 80 nm, 20/2 nm, and 1.6 μm, respectively. Compositional-contrast SEM using backscattered electron detection (BSED), shown in Figure 1(b), reveals four distinct regions across the heterostructure identified (from top to bottom) as the GaN channel, super lattice 1 (SL 1), and super lattice 2 (SL 2). The corresponding thicknesses of GaN, SL 1, and the SL 2 channel are approximately 1.6 μm, 2.2 μm, and 1.2 μm, respectively. TEM analysis in figure 1(c) and 1(d) also confirms that both underlying buffer layers consist of engineered AlGaN/AlGaN superlattices designed for strain relaxation and dislocation suppression. Buffer layer 1 features a superlattice period of 24.5 nm, comprising $Al_xGa_{1-x}N$ sublayers of 22.5 nm and 2 nm. Buffer layer 2 exhibits a slightly larger period of 25.5 nm, consisting of 20.5 nm and 5 nm sublayers. A gradual reduction in Al composition toward the channel region is also observed *via* EDS measurement. These superlattice configurations improve *crystalline* quality by mitigating threading dislocation propagation into the GaN channel[33, 34].

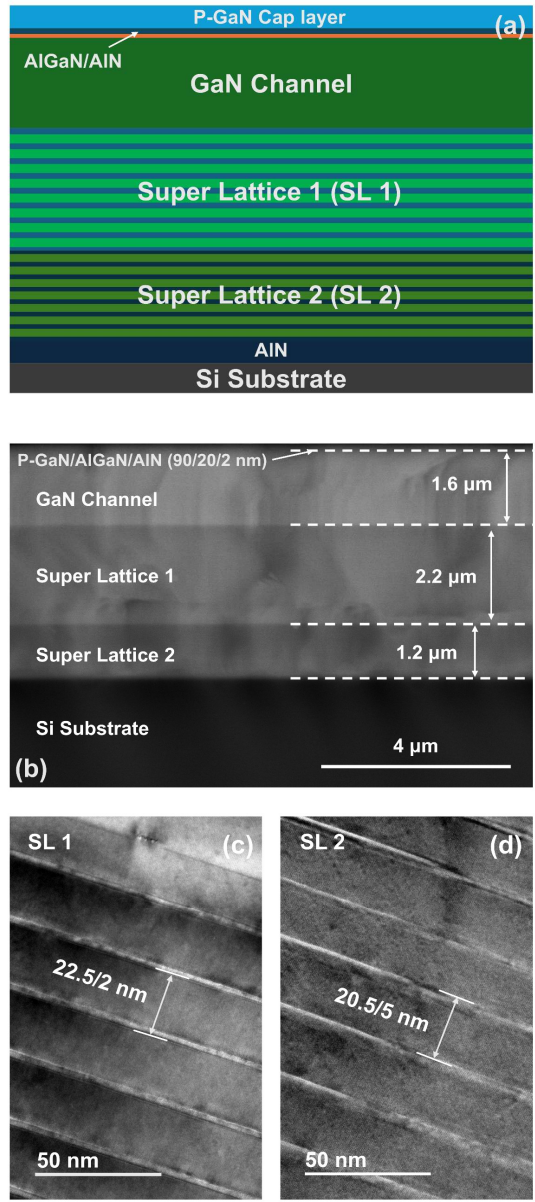

**FIG. 1.** (a) Schematic illustration of the epitaxial layer structure of the GaN HEMT heterostructure grown on a Si (111) substrate. (b) Cross-sectional SEM image of the full HEMT structure. (c, d) High-resolution TEM images of the two superlattice buffer regions of SL 1 and SL 2.

Figure 2(a) shows *2θ–ω* XRD patterns of the GaN (0002) reflection and its surrounding satellite peaks at various temperatures, measured *via* HT-XRD up to 1000 °C. The main diffraction peak at approximately 34.6° corresponds to the GaN channel layer, while the periodic satellite peaks on either side of the GaN peak, originates from the two AlGaN/GaN superlattice buffer layers with the main peak at 34.9° and 35.2°, respectively. As the temperature rises, all diffraction peaks shifted slightly toward lower 2θ angles. Figure 2(b) extracts the peak position changing of Si substrate, GaN, SL 1, and SL 2 peak at room temperature before and after the heating process, as well as the changes from 500 °C to 1000 °C. According to Bragg's law, the relationship between diffraction angle '$\theta$' and the vertical lattice plane distance '$d$' is given by[35]:

$$n\lambda = 2d\sin(\theta) \tag{1}$$

where '$\lambda$' is the X-ray wavelength. Thus, a decrease in $\theta$ implies an increase in the vertical lattice constant $c$, since $\lambda$ remains constant. This observation aligns with the lattice thermal-expansion theory.

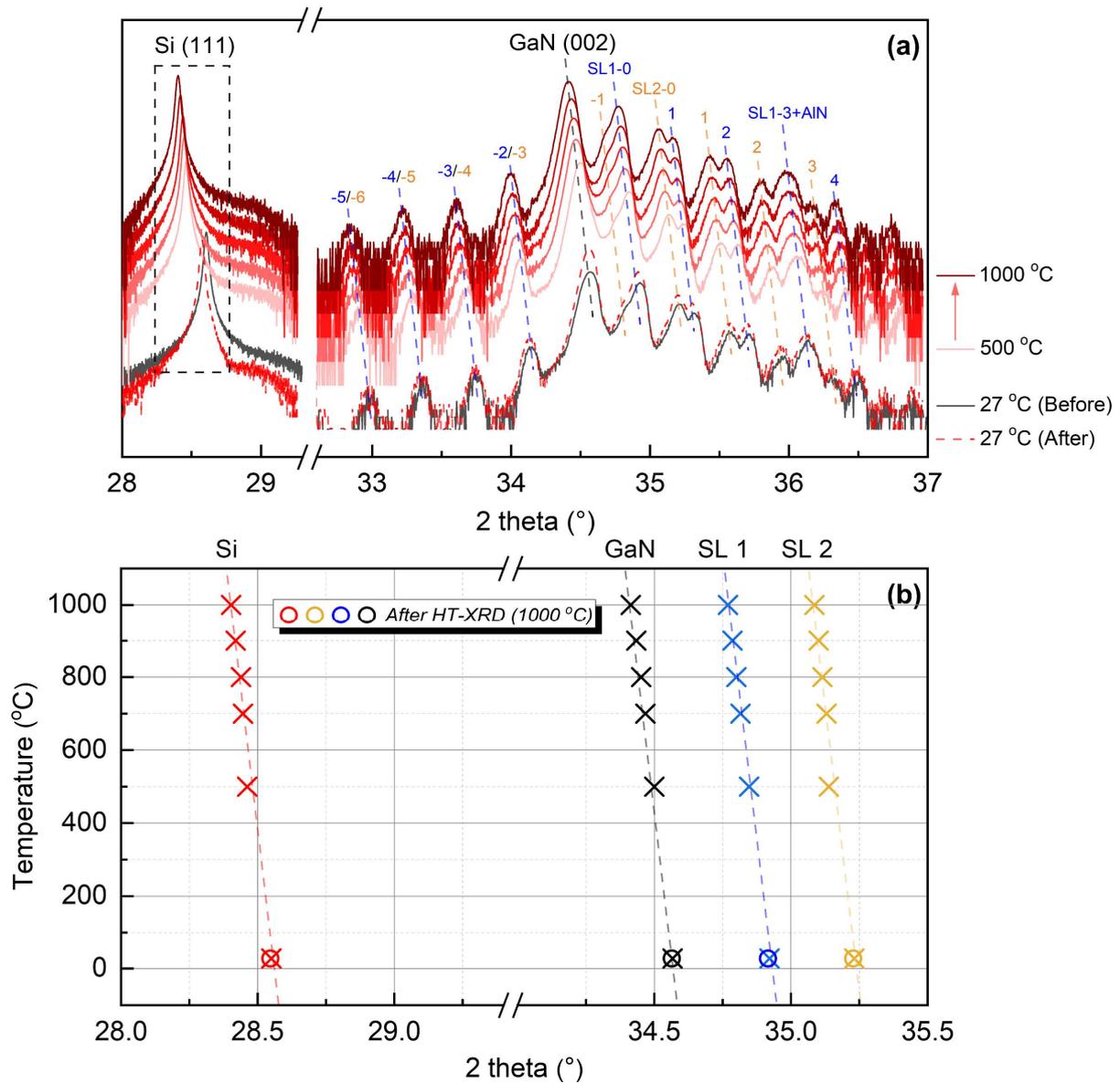

**FIG. 2.** (a) High-temperature X-ray diffraction (HT-XRD) *2θ–ω* scans of the GaN HEMT heterostructure recorded at various temperatures from 27 °C to 1000 °C and subsequently cooling back to 27 °C. (b) Extracted 2θ peak positions of Si, GaN, SL 1, and SL 2 as a function of temperature.

To further investigate the lattice evolution under elevated temperatures, both symmetric and asymmetric reciprocal space mapping (RSM) measurements were conducted during the HT-XRD process up to 1000 °C. Figures 3(a) and 3(b) present the symmetric and asymmetric RSMs, respectively, from 27 °C to 1000 °C and cooled back to 27 °C. In the symmetric RSM (Fig. 3(a)),

Si and GaN reflections correspond to the (111) and (0002) planes, respectively, whereas in the asymmetric RSM (Fig. 3(b)), they are identified as (224) and (10-15) reflections, respectively. Figure 3(c) and 3(d) shows the zoom in symmetric and asymmetric RSM image around the GaN peak, respectively. The difference between superlattice main peak and satellite peaks is unclear in the symmetric scan, while show interlaced arrangement along the $q_x$ axis in the asymmetric scan. Because the AlGaN/GaN superlattice is periodically modulated only along the growth direction, its Fourier components add replicas of each Bragg reflection shifted by integer multiples exclusively along $q_z$:

$$Q = \frac{2\pi}{\Lambda} \quad (1),$$

which $\Lambda$ is the period of superlattice, producing a vertically aligned series of satellite peaks in the RSM. Thus, the vertically distributed peaks can be distinguished as the satellite peaks, while the strong peaks can be distinguished as the GaN, SL 1, SL 2 with different $q_x$ value.

The RSMs capture diffraction signals from the Si substrate, the GaN channel, and the well-defined superlattice (SL) satellite peaks at all temperatures. No broadening or distortion of the GaN and SL peaks were observed throughout the thermal cycling process, indicating thermal stability of the overall heterostructure. At 27 °C, the different lateral alignment position of GaN and Si peak in symmetric scan represents the relaxed lattice condition in GaN channel layer. As the temperature increases, the $q_z$ positions of the GaN and Si diffraction peak systematically shifted toward lower values, reflecting the expected thermal expansion of the out-of-plane lattice spacing ($q_z \propto 1/c$). Meanwhile, the lateral peak alignment between GaN and Si in the asymmetric RSM remained unchanged over the entire thermal cycle, suggesting minimal in-plane relaxation and stable strain state in the buffer and channel regions. The SL satellite peaks remain sharp and periodically distributed, further verifying the preservation of superlattice periodicity and interface integrity even up to 1000 °C.

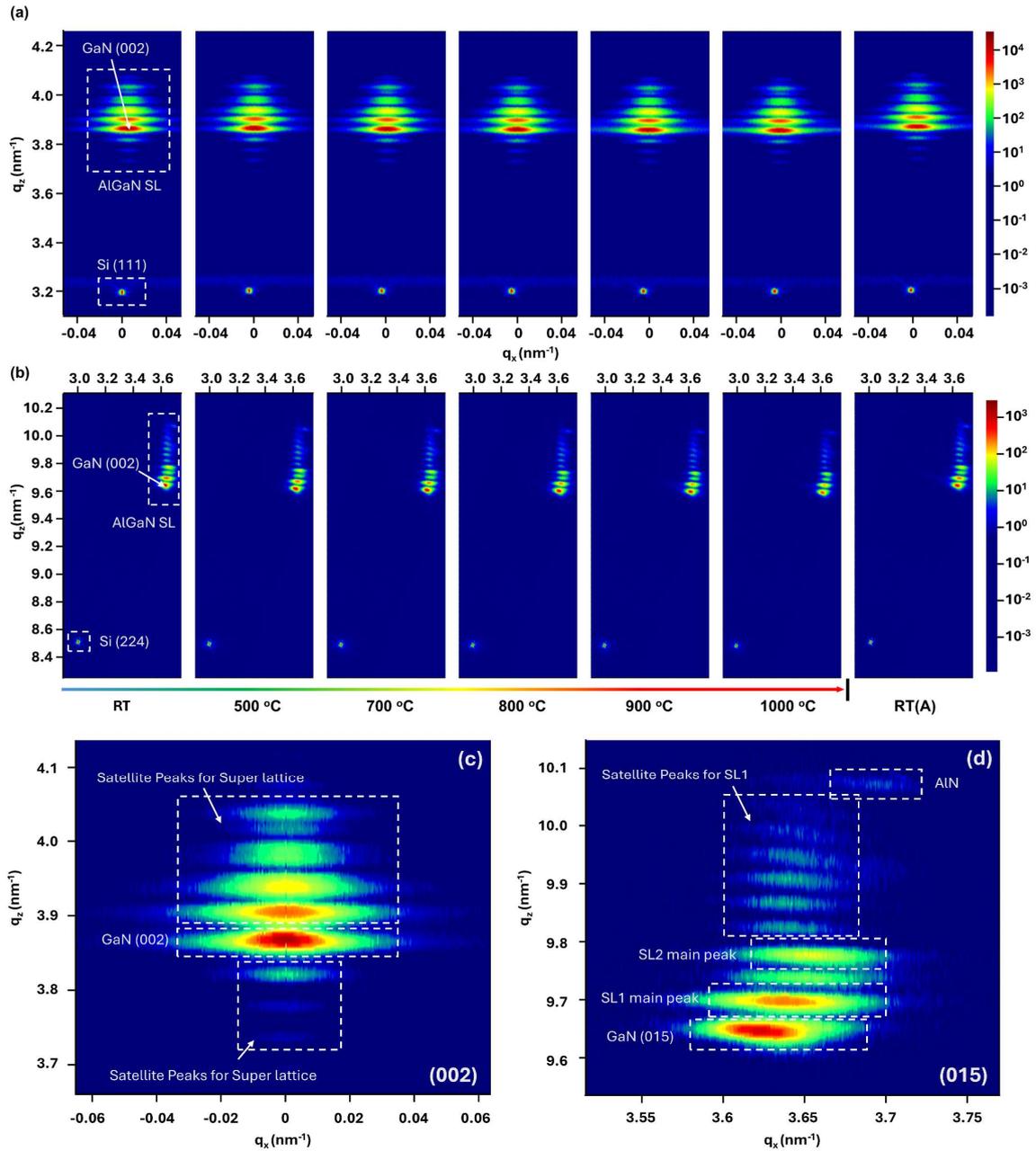

**FIG. 3.** (a) Symmetric (top) and (b) asymmetric (bottom) *in situ* reciprocal space maps (RSMs) of the GaN HEMT heterostructure measured at different temperatures. From 27 °C to 1000 °C and after cooling. (c) Enlarged RSM around the GaN(002) peak. (d) Enlarged asymmetric RSM around GaN(015).

Based on Bragg's law, the variation of '*c*' and the corresponding out-of-plane strain were calculated through *2θ* positions in *2θ-ω* scans. Figure 4(a) shows that the *c*-lattice constantly changed from 27 °C to 1000 °C. As seen in Figure 4, the lattice constant $c$ expanded linearly from 5.195 Å to 5.208 Å as the temperature increased from 27 °C to 1000 °C. To evaluate this thermal expansion, a predicted curve of GaN $c$-lattice versus temperature from the multi-frequency Einstein model (ME model) is included in the plot (Figs. 4(a-b)). The thermal expansion coefficient α changing can be fitted through this model as[36, 37]:

$$\alpha = \sum_{i=1}^{n} X_i \frac{\left(\frac{\theta_i}{T}\right)^2 \exp\left(\frac{\theta_i}{T}\right)}{\left[\exp\left(\frac{\theta_i}{T}\right)-1\right]^2} \tag{3},$$

where $X_i$ and $\theta_i$ are both fitting parameters. With the fitted thermal expansion coefficient α, the expanded lattice constant can be calculated via simple integration as:

$$a(T) = a(T_{ref}) \exp\left(\int_{T_{ref}}^{T} \alpha_a(T') \, dT'\right) \tag{4},$$

$$c(T) = c(T_{ref}) \exp\left(\int_{T_{ref}}^{T} \alpha_c(T') \, dT'\right) \tag{5},$$

where $T_{ref}$ is the reference temperature. In comparison, the measured lattice constant $c$ matches well with the predicted curve, demonstrating that the lattice is expanding freely along the vertical direction. By comparing the expanding vertical lattice with the theoretical value of 5.185 Å, the out-of-plane strain was determined to evolve with increasing temperature, reaching 0.44 % at 1000 °C. Upon cooling back to 27 °C, the lattice relaxed to 0.017 %, suggesting a minor, irreversible lattice distortion after the thermal cycle.

To evaluate the in-plane strain evolution of GaN during the HT-XRD process, the lateral lattice constant ($a$) was calculated from the $q_x$ value of the GaN (015) reflection in the asymmetric RSM (Fig. 3(b)). The relationship between the reciprocal-lattice coordinate $q_x$ and the in-plane lattice parameter $a$ can be expressed by Eq. 2[38]:

$$q_x = \frac{1}{a}\sqrt{\frac{4}{3}(h^2+hk+k^2)} \tag{6},$$

After inserting the corresponding indices for the (015) plane, the equation simplifies to a direct conversion between $q_x$ and the $a$-lattice constant of GaN to Eq. 3[38]:

$$a = \frac{2}{\sqrt{3}q_x} \tag{7},$$

Figure 4(b) presents the evolution of the calculated in-plane lattice constant ($a$) of the GaN channel, Si substrate, SL 1 and SL 2 as a function of temperature from 27 °C to 1000 °C, along with calculated prediction curves. Unlike the $c$-axis lattice constant, the $a$-lattice expansion does not follow a perfectly linear trend, and the expansion rate gradually decreases with increasing temperatures. The experimental data exhibits a slightly slower slope than the theoretical GaN curve. Initially, the $a$ lattice constant of GaN layer shows the value of 3.185 Å which is smaller than the theoretical value of 3.189 Å. The compressive strain of 0.12% in GaN comes from the smaller lateral lattice constant of $Al_xGa_{1-x}N$ superlattice buffer layer. As the temperature increases, the compressive effect from the AlGaN superlattice layers restricts the thermal expansion of GaN channel layer. This interplay leads to a situation in which the GaN in-plane strain becomes smaller than both the intrinsic GaN and Si thermal-expansion predictions, because the AlGaN layers effectively suppress further lateral elongation of $a$. Consequently, the measured thermal-expansion rate of the GaN film is slower than theoretical prediction for free GaN, reflecting the constrained strain evolution within the multilayer heterostructure. To verify the thermal expansion behavior, the predicted temperature-dependent relative lattice change percentage of Si and GaN was compared with the experimentally measured percentage of GaN, Si substrate, SL 1 and SL 2, as shown in Figure 4(c)[39]. A clear trend is observed that all four layers from the Si substrate through SL1, SL2, and up to the GaN layer, exhibiting nearly the same thermal-expansion ratio up to 900 °C, as further illustrated in Fig. 4(e). In the schematic, the atomic positions of each layer align along nearly the same straight line passing through the center at a given temperature, indicating that their in-plane expansion proceeds coherently and proportionally. However, at 1000 °C, Fig. 4(c) reveals a pronounced divergence in the percentage change among the layers. This misalignment, also sketched in the schematic, signifies the onset of substantial strain accumulation due to thermal-expansion mismatch. At this elevated temperature, the coherent expansion breaks down, increasing the dislocation generation and defect formation within the superlattice and at the interfaces.

After the sample was cooled back to 27 °C, the $a$-lattice constant did not return to its initial value (3.185 Å), but instead became slightly smaller (3.117 Å), corresponding to a residual compressive with the percentage of approximately 0.3%. This irreversible lattice contraction is attributed to the rapid cooling process (~15 min) compared to the much longer high-temperature XRD measurement duration (~3 h). During the rapid temperature drop, the lattice contracts quickly shrink from their thermally expanded state, but the restricted relaxation dynamics limit complete strain recovery. As a result, partial stress reversal occurs, leaving a compressive residual strain in the GaN lattice after the HT-XRD cycle.

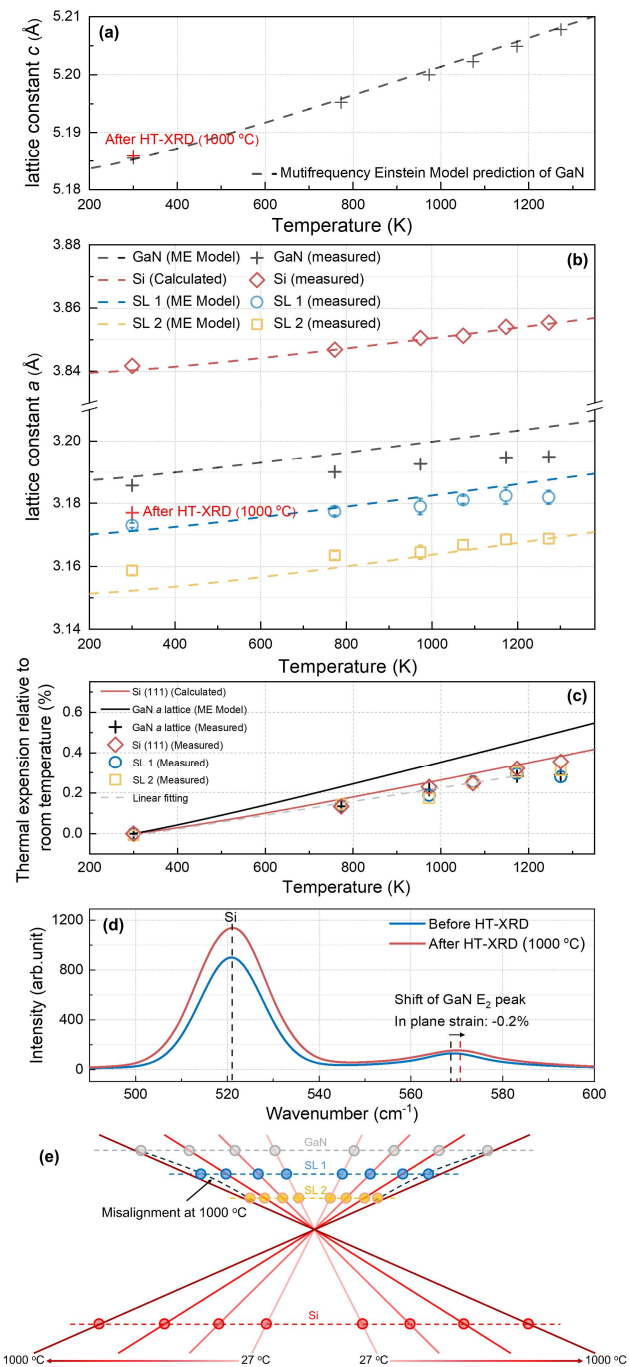

**Figure 4.** Temperature-dependent variation of the (a) c-lattice constant and (b) a-lattice constant derived from the symmetric RSM with the prediction of the multi-frequency Einstein thermal-

expansion model (ME model). (c) Comparison of the predicted in-plane strain evolution of GaN and Si with temperature, together with the experimentally measured in-plane strain of the GaN layer. (d) Raman spectra of the GaN HEMT sample measured before and after the 1000 °C HT-XRD process. (e) Schematic illustration of the thermal-expansion behavior of the Si substrate, SL2, SL1, and GaN layer from room temperature to 1000 °C.

To further verify the strain evolution induced by the HT-XRD process, Raman spectroscopy was performed on the GaN HEMT sample before and after the high-temperature annealing[40], as seen in Figure 4(d). The prominent Si (520 cm$^{-1}$) peak remains unchanged, while the E$_2$ (high) phonon mode of GaN shifts slightly toward higher wavenumber (blue shift) after the 1000 °C HT-XRD cycle, indicating the development of compressive strain within the GaN layer. The in-plane strain ($\varepsilon_\parallel$) can be quantitatively estimated from the phonon shift ($\Delta\omega$) of the E$_2$ (high) mode using the phonon-strain relationship in Eq. 4[41]:

$$\varepsilon_\parallel = \frac{\Delta\omega}{K} \quad (8),$$

where $K$ is the phonon deformation potential of the GaN E$_2$ (high) mode, typically reported as -780 cm$^{-1}$ per unit strain[42]. Based on the measured blue shift of approximately 1.5 cm$^{-1}$ after the HT-XRD process, the corresponding in-plane strain was calculated to be around -0.2%, indicating a small compressive strain compared with the as-grown sample. This strain magnitude is in good agreement with the value obtained from the asymmetric RSM analysis of the $a$-lattice constant (-0.3%), confirming that the rapid cooling following the 1000 °C exposure led to partial strain reversal in the GaN layer. The combined XRD and Raman results consistently demonstrate that the HT-XRD cycle did not significantly degrade the crystalline quality but introduced thermally induced compressive strain due to the non-equilibrium lattice contraction during cooling.

To further evaluate the crystalline quality evolution induced by the high-temperature process, rocking curve (RC) measurements were performed on the GaN HEMT sample before and after the HT-XRD experiment. Figure 5(a) compares the RC profiles of both the symmetric (002) and asymmetric (204) reflections. After the high-temperature cycle, the full width at half maximum (FWHM) of both reflections exhibited a slight increase, indicating a moderate degradation in crystal structure. Specifically, the FWHM of the (002) axis broadened from 0.1568° to 0.1772°, while the FWHM of the (204) axis increased from 0.2153° to 0.2282°. The dislocation densities were subsequently calculated using the following well-established relationship between FWHM and dislocation density[40, 43]:

$$\rho = \frac{\beta^2}{4.35\, b^2} \tag{9}$$

where $\rho$ is the dislocation density, $\beta$ is the FWHM (in radians), and $b$ is the magnitude of the relevant Burgers vector ($b = c = 0.5185$ nm for screw dislocations and $b = a = 0.3189$ nm for edge dislocations). Based on these calculations, the screw dislocation density increases from $6.4 \times 10^{10}$ cm$^{-2}$ to $8.7 \times 10^{10}$ cm$^{-2}$ (28 % increase), while the edge dislocation density increases from $3.17 \times 10^{11}$ cm$^{-2}$ to $3.56 \times 10^{11}$ cm$^{-2}$ (12 % increase), as summarized in Figure 5(b). The higher density of edge dislocations relative to screw dislocations is consistent with the previous RSM and Raman analyses, confirming that the lateral (in-plane) lattice remains more constrained due to residual mismatch strain from the AlGaN/AlGaN buffer and the Si (111) substrate, while the vertical lattice direction is relatively relaxed. The slightly larger relative increase in screw dislocation density, however, suggests that partial strain redistribution occurred along the vertical direction during the high-temperature process. Overall, the comparable order of magnitude for both dislocation types indicates that the crystal quality degradation is modest and elastic, without significant generation of new extended defects.

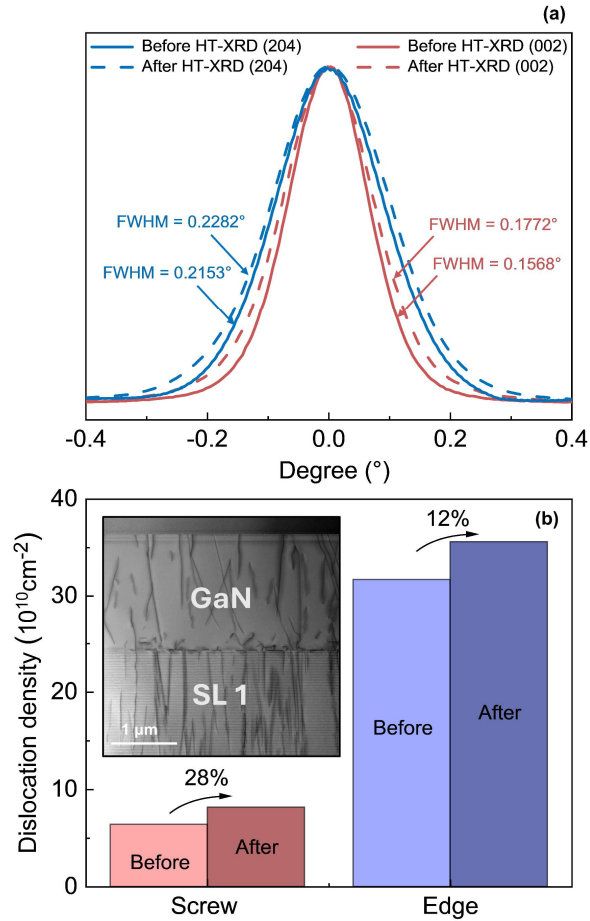

**FIG. 5.** (a) Rocking curve (RC) profiles of the GaN (002) symmetric and (204) asymmetric reflections measured before and after the 1000 °C HT-XRD process. (b) Comparison of the calculated screw and edge dislocation densities before and after HT-XRD.

In summary, a comprehensive structural investigation of a GaN-on-Si E-mode HEMT heterostructure was conducted using *in situ* HT-XRD and complementary Raman spectroscopy to elucidate the lattice and defect evolution under extreme thermal conditions. The GaN channel and AlGaN/AlGaN superlattice buffers retained distinct diffraction and satellite features up to 1000 °C, confirming high thermal stability and interface robustness. The vertical lattice expansion closely followed theoretical predictions, while the in-plane expansion exhibited deviation due to residual tensile strain originating from the Si substrate and buffer mismatch. After thermal cycling, a residual compressive strain of ~0.3% was identified, consistent with the ~0.2% strain derived from Raman peak shifts. Rocking-curve analysis further revealed a moderate increase in dislocation densities, suggesting minor strain redistribution but no substantial crystalline degradation. Overall, the results indicate that the GaN-on-Si HEMT structure maintains excellent crystalline integrity and thermal reliability up to a temperature of 1000 °C. The combined *in situ* HT-XRD and Raman approach provides valuable insights into the strain relaxation mechanisms and defect dynamics in GaN heterostructures, offering practical guidance for optimizing buffer design and enhancing the thermal robustness of next-generation high-temperature GaN power devices.

## AUTHOR DECLARATIONS

### Conflict of Interest

The authors have no conflicts to disclose.

### Author Contributions

**Botong Li:** Conceptualization (equal); Data curation (equal); Formal analysis (equal); Investigation (equal); Methodology (equal); Visualization (equal); Writing - original draft (equal); Writing - review & editing (equal). **Bobby George Duersch:** Formal analysis (equal); Investigation (equal); Methodology (equal); Resources (equal); Validation (equal); Visualization (equal); Writing - original draft (equal); Writing - review & editing (equal). **Hunter D. Ellis:** Investigation (equal); Methodology (supporting); Writing - review & editing (equal). **Imteaz Rahaman:** Formal analysis (supporting); Investigation (supporting); Methodology (supporting); Writing - original draft (equal); Writing - review & editing (supporting). **Aidan J. Belanger:** Formal analysis (equal); Investigation (equal); Methodology (equal); Software (equal); Writing - review & editing (equal). **Zlatan Aksamija:** Formal analysis (supporting); Investigation (equal); Methodology (supporting); Resources (equal); Software (equal); Writing - review & editing (equal). **Brian Roy Van Devener:** Investigation (equal); Methodology (supporting); Validation (equal); Visualization (equal). **Kathy Anderson:** Methodology (supporting); Project administration (equal); Resources (equal). **Kai Fu:** Conceptualization (lead); Investigation (supporting); Methodology (lead); Project administration (lead); Resources (lead); Supervision (lead); Validation (equal); Writing – original draft (lead); Writing – review & editing (lead).

**ACKNOWLEDGEMENT**

The authors acknowledge the support from the University of Utah start-up fund and Research Incentive Seed Grant by the Price College of Engineering and the Vice President for Research (VPR) Office. This work made use of the Nanofab EMSAL shared facilities of the Micron Technology Foundation Inc. Microscopy Suite, sponsored by the John and Marcia Price College of Engineering, Health Sciences Center, Office of the Vice President for Research. In addition, it utilized the University of Utah Nanofab shared facilities, which are supported in part by the MRSEC Program of the NSF under Award No. DMR-112125. Acquisition of the Bruker D8 Discover system was made possible by the Air Force Office of Scientific Research under project number FA9550-21-1-0293.

**DATA AVAILABILITY**

The data that supports the findings of this study are available from the corresponding authors upon reasonable request.